\documentclass[conference]{IEEEtran}
\IEEEoverridecommandlockouts
\usepackage{cite}
\usepackage{amsmath}
\usepackage{amsmath,amssymb,amsfonts}
\usepackage{pifont}
\usepackage{algorithm}
\usepackage{algorithmic}
\usepackage{graphicx}
\usepackage{textcomp}
\usepackage{xcolor}
\usepackage{hyperref}
\usepackage{listings}
\usepackage{enumitem}
\usepackage{balance}

\usepackage{booktabs}
\usepackage[group-separator={,},group-minimum-digits=4]{siunitx}
\usepackage{xcolor}
\usepackage{tikz}
\definecolor{codegreen}{rgb}{0,0.6,0}
\definecolor{codegray}{rgb}{0.5,0.5,0.5}
\definecolor{codepurple}{rgb}{0.58,0,0.82}
\definecolor{backcolour}{rgb}{0.95,0.95,0.92}
\newcommand{\cmark}{\ding{51}}
\newcommand{\xmark}{\ding{55}}

\makeatletter
  \def\sectionautorefname{\S\@gobble}
  \def\subsectionautorefname{\S\@gobble}
  \def\subsubsectionautorefname{\S\@gobble}

  \newcommand{\removelatexerror}{\let\@latex@error\@gobble}
\makeatother

\newif\iffinal
\iffinal
  \newcommand{\zhuozhao}[1]{}
    \newcommand{\kyle}[1]{}
      \newcommand{\yifei}[1]{}
      \newcommand{\ryan}[1]{}
    \newcommand{\ian}[1]{}

\else
  \newcommand{\zhuozhao}[1]{{\textcolor{red}{ ZZ: #1 }}}
    \newcommand{\kyle}[1]{{\textcolor{red}{ Kyle: #1 }}}
      \newcommand{\yifei}[1]{{\textcolor{magenta}{ Yifei: #1 }}}
      
      \newcommand{\ryan}[1]{{\textcolor{magenta}{ Ryan: #1 }}}
    \newcommand{\ian}[1]{{\textcolor{blue}{ Ian: #1 }}}

\fi
\newcommand\mytt[1]{{\small\texttt{#1}}}

\newcommand{\name}{UniFaaS}
\newcommand{\funcxname}{\textit{funcX}}
\newcommand{\parslname}{\textit{Parsl}}
\newcommand{\taiyi}{Taiyi}
\newcommand{\qiming}{Qiming}
\newcommand{\csecluster}{Dept. cluster}
\newcommand{\lab}{Lab cluster}
\newcommand{\pc}{Workstation}
\newcommand{\decorator}{function}
\lstdefinestyle{myPython}{
    language=python,
    morekeywords={Config, Executor, load, scheduling_strategy, @python_app, GlobusFile, @function, max_transfer_retries, 
    result_polling_interval, enable_task_stealing, enable_probe_file,file_transfer_type,executors},
    backgroundcolor=\color{backcolour},   
    commentstyle=\color{codegreen},
    keywordstyle=\color{magenta},
    numberstyle=\tiny\color{codegray},
    stringstyle=\color{codepurple},
    basicstyle=\fontfamily{pbk}\footnotesize, % or pcr for Courier
    breakatwhitespace=false,         
    breaklines=true,                 
    captionpos=b,                    
    keepspaces=true,                 
    numbers=left,                    
    numbersep=5pt,                  
    showspaces=false,                
    showstringspaces=false,
    showtabs=false,                  
    tabsize=2
}

\def\BibTeX{{\rm B\kern-.05em{\sc i\kern-.025em b}\kern-.08em
    T\kern-.1667em\lower.7ex\hbox{E}\kern-.125emX}}

\begin{document}

\DeclareRobustCommand*{\IEEEauthorrefmark}[1]{%
  \raisebox{0pt}[0pt][0pt]{\textsuperscript{\footnotesize #1}}%
}

\title{\name{}: Programming across Distributed Cyberinfrastructure with Federated Function Serving

\thanks{$^*$Zhuozhao Li is the corresponding author. Email: lizz@sustech.edu.cn.}
}

\author{
    \IEEEauthorblockN{Yifei Li\IEEEauthorrefmark{1}, Ryan Chard\IEEEauthorrefmark{2}, Yadu Babuji\IEEEauthorrefmark{3}%\IEEEauthorrefmark{3}
    , Kyle Chard\IEEEauthorrefmark{3,}\IEEEauthorrefmark{2}, Ian Foster\IEEEauthorrefmark{2,}\IEEEauthorrefmark{3}, Zhuozhao Li\IEEEauthorrefmark{1,}$^*$}
    \IEEEauthorblockA{\IEEEauthorrefmark{1}Dept. of Computer Science and Engineering, Southern University of Science and Technology, Shenzhen, China}
    \IEEEauthorblockA{\IEEEauthorrefmark{2}Data Science and Learning Division, Argonne National Laboratory, Lemont, IL, USA}
    \IEEEauthorblockA{\IEEEauthorrefmark{3}Dept. of Computer Science, University of Chicago, Chicago, IL, USA}
    
}

\maketitle
%\IEEEpeerreviewmaketitle
%\thispagestyle{plain}
%\pagestyle{plain}
\setcounter{page}{1}

\begin{abstract}
Modern scientific applications are increasingly decomposable into individual functions that may be deployed across distributed and diverse cyberinfrastructure such as supercomputers, clouds, and accelerators. 
Such applications call for new approaches to programming, distributed execution, and function-level management.
We present \name{}, a parallel programming framework that relies on a federated function-as-a-service (FaaS) model to enable composition of distributed, scalable, and high-performance scientific workflows, and to support fine-grained function-level management. 
\name{} provides a unified programming interface to compose dynamic task graphs with transparent wide-area data management.
\name{} exploits an observe-predict-decide approach to efficiently map workflow tasks to target heterogeneous and dynamic resources.
We propose a dynamic heterogeneity-aware scheduling algorithm that employs a delay mechanism and a re-scheduling mechanism to accommodate dynamic resource capacity.

Our experiments show that \name{} can efficiently execute workflows across %a sufficient number (e.g., 16) of 
computing resources with minimal scheduling overhead.
% Through a drug screen workflow, 
We show that \name{} can improve the performance of a real-world drug screening workflow by as much as 22.99\% when employing an additional 19.48\% of resources and a montage workflow by 54.41\% when employing an additional 47.83\% of resources across multiple distributed clusters, in contrast to using a single cluster.

\end{abstract}

\begin{IEEEkeywords}
federated cyberinfrastructure, federated function serving
\end{IEEEkeywords}

\section{Introduction}\label{sec:intro}
The rapid adoption of hardware accelerators and exponential increases in data volumes have spurred a major transformation in the nature of programming. 
Developers increasingly decompose previously monolithic workflows into many individual, often lightweight tasks~\cite{shaffer2021lightweight}. 
These tasks may be individual functions or executables developed in different programming languages, with %differing sizes and 
varied resource requirements---from single-core functions and multi-core simulation codes, to accelerator-based computations. 
When workflows are decomposed in this way, it is then feasible, and indeed often desirable, to execute different tasks in different locations: for example, where accelerators or resources are available, where software is installed, near data, or with the best cost-performance efficiency. 

Unfortunately, such fine-grain workflows are not well supported by current research cyberinfrastructure (CI) that is managed by batch schedulers designed for scheduling large batch jobs and typically have long (and unpredictable) queue times. As researchers often have access to a \emph{resource pool} containing several computing resources (e.g., institutional clusters, supercomputers, and accelerators), one may want to exploit these resources concurrently to amortize queue times (though perhaps at the cost of additional data transfers) for faster scientific analyses. For example, it may take only minutes to transfer gigabytes and hours to transfer terabytes~\cite{liu2017explaining}, but days to obtain hundreds of nodes on an oversubscribed supercomputer.
%Due to the need to run functions at different locations and for faster analyses, 

As a result, researchers increasingly use multiple resources \emph{together} across federated CI (e.g., supercomputers, clouds, accelerators, and local compute) to run a single workflow---a landscape referred to as \emph{cross-facility computing}~\cite{da2023workflows}.
Such a trend towards fine-grained and distributed scientific workflows drives new system requirements. 
To name just a few (more are described in \autoref{sec:requirements}): 
1) \emph{programmability} to allow users to construct programs able to execute on diverse and distributed CI with minimal barriers; 
%2) transparently managing \emph{data dependencies} among tasks, as moving data in the computing continuum is challenging; 
2) automatic \emph{fine-grained} task-level management in terms of resource provisioning and heterogeneous environment management; 
and 3) enabling \emph{flexible} task execution on dynamic resources as the resource availability across federated CI may change at runtime.

Myriad systems have been developed to address the needs of users writing and deploying workflows~\cite{alsaadi2021radical}. 
While these systems address some of the requirements above, none address the entire set of requirements. 
% differ in various aspects. 
For example, many workflow systems~\cite{di2017nextflow,wilde2011swift} rely on domain-specific languages to support workflow composition. Python-based computing frameworks such as Ray~\cite{moritz2018ray}, Parsl~\cite{parsl}, and Dask~\cite{rocklin2015dask} support development of parallel programs on a single computer but cannot easily adapt to federated CI. Pegasus~\cite{deelman2015pegasus} %and DAGMan~\cite{thain2005distributed}
leverages a static configuration-based model to compose scientific workflows as directed-acyclic graphs (DAGs) but requires deployment of HTCondor~\cite{thain2005distributed} which is rarely supported on HPC systems and cannot support fine-grained task-level management.

While in principle one can develop modern workflows by writing batch scripts or combining various workflow systems for federated CI, this approach in practice not only imposes a significant development and management burden (e.g., fault tolerance, data movement, and resource management) on researchers, but is also not flexible, scalable, or portable across computers, reducing efficiency and reproducibility that are crucial for science~\cite{madduri2019reproducible}.

We present \name{}, a general-purpose, parallel programming framework that adapts a federated function-as-a-service (FaaS) model to enable developers to compose distributed, scalable, and high-performance scientific workflows that span federated CI. The FaaS model, first implemented by commercial cloud providers~\cite{jonas2019cloud}, can benefit modern scientific applications from several perspectives.
FaaS reduces the burden of managing computing resources and instead exposes a function-based API that allows users to manage and schedule scientific workflows at a fine-grained function level. Further, it enables the use of containers to both simplify the management of complicated environments and reduce portability challenges. Finally, FaaS, by definition, abstracts the resource pool and thus enables resources to be dynamically added (or removed) during execution.

We implement \name{} upon \funcxname{}~\cite{li2022funcx,chard2020funcx}, a \emph{federated} FaaS platform that extends the FaaS model and enables users to invoke functions on arbitrary resources. 
\name{} provides a \emph{unified} programming interface to express task parallelism and compose dynamic dependency graphs, which can be further deployed across distributed resources seamlessly. 
\name{} implements a data manager to \emph{transparently} manage data transfers across computers on behalf of users, using widely-used transfer mechanisms such as Globus~\cite{chard2014efficient} and rsync.

It is non-trivial to achieve high performance in cross-facility computing, since the resources across federated CI are heterogeneous and each resource may have dynamic capacity during execution due to scheduled downtimes and use by others.

In this paper, we explore an observe-predict-decide approach to improve the performance: \name{} monitors the key characteristics (e.g., input data sizes and environments) of tasks on different computers and predicts the task performance (e.g., transfer and execution time) using common performance models.
We propose a dynamic heterogeneity-aware scheduling algorithm that employs a delay mechanism and a re-scheduling mechanism to accommodate the workflow and resource dynamics. 
\name{} supports elasticity---it can automatically scale various resources based on workflow characteristics. 
Furthermore, the modular design of \name{} allows users to easily plug in any appropriate schedulers or data transfer mechanisms for their workflows.

To the best of our knowledge, \name{} is the first framework that adapts the convenient federated FaaS model to enable monolithic workflows to be broken into small schedulable function units that can be flexibly executed across a resilient, federated resource pool. In this paper, we explore the feasibility of adapting the FaaS paradigm and aim to design an extensible system to address the unique challenges of programmability, scheduling, and data management (more in \autoref{sec:requirements}) in cross-facility computing. %We further explore programming and scheduling approaches under the unique federated FaaS model.
Together, these features enable productive parallel programming and simple, yet performant, execution management on federated CI.

The contributions of our work are as follows: 

\begin{itemize}[leftmargin=0.25in]
    \item We design \name{}, a general-purpose parallel programming framework that simplifies workflow structuring and enables fine-grained function-level management of workflows in federated environments. 
    %computing continuum. 
    \item We propose a dynamic heterogeneity-aware scheduler and a re-scheduling mechanism to handle heterogeneous and dynamic computing environments.
    \item Our evaluation shows that \name{} can deploy workflows across up to 16 computing resources with high performance. 
    \item We discuss lessons learned when using \name{} in real use cases and identify areas for further improvement. 
\end{itemize}

The rest of this paper is as follows.
\autoref{sec:requirements} %describes an example use case and 
presents general requirements for modern scientific workflows.
\autoref{sec:programming} presents the programming model of \name{}.
\autoref{sec:arch} describes the \name{} system architecture, optimizations, and implementations.
% \autoref{sec:implementation} presents the implementations and optimizations.
\autoref{sec:evaluation} evaluates \name{}'s performance. 
\autoref{sec:discussion} discusses the lessons learned with \name{}.
\autoref{sec:related} discusses related work.
\autoref{sec:conclusion} concludes the paper with future remarks.

\section{Motivation and Background}\label{sec:requirements}
In this section, we highlight the key requirements for modern scientific workflows deployed across federated CI by describing a real-world use case. 
We note that these characteristics are shared by many use cases, for example in modern AI-driven simulations and data-driven workflows~\cite{thomert2019towards,abdelBaky2017computing,rosendo2022distributed,clyde2022high,pauloski2023accelerating,ward2021colmena,chard2019dlhub,li2021dlhub,skluzacek2021serverless,woodard2020real}.
Further, recent work has shown that such distribution can also improve energy consumption of workflows~\cite{CAWS}. 

The COVID-19 pandemic highlighted the need for discovery of effective therapeutics among a near-infinite search space of small molecules. 
One common way to accelerate the design and development of antiviral treatments is to use machine learning (ML) models to computationally screen small molecules rapidly, much faster than is possible with wetlab studies.
A typical drug screening pipeline~\cite{saadi2021impeccable} involves multiple computational stages, such as molecular simulations, feature computations, fingerprinting, etc., each with distinct computational needs. %docking calculations, and similarity search,
Each stage relies on different toolkits and methods (e.g., simulation, machine learning), requires different types of resources (e.g., CPUs, accelerators) and different amounts of those resources, and has diverse task features (e.g., parallelism, dependency, duration, input size).
Computers also vary significantly in characteristics, for example, some may be powerful but have long queue times while others may have fewer resources but are immediately available.
An application that needs, for example, to screen millions of molecules quickly must be able to adapt to these different tasks and resource characteristics.

These modern use cases motivate the need for deployment in federated environments and exhibit the following key requirements.

%\begin{itemize}
\begin{enumerate}[leftmargin=0.2in]
\item \textbf{Fine-grained management}: workflows usually consist of tasks with diverse requirements on resources, environments, software dependencies, data locations, etc., and thus require fine-grained task-level management automatically to reduce the burden on users. 
\item \textbf{Portability}: scientific applications require portability in different dimensions: first, a workflow may run on various computing resources with different environments for reproducing and sharing~\cite{saadi2021impeccable}; second, a stage may be portable to different computations, e.g., different ML models or different methods.
\item \textbf{Elasticity}: workflows may have different resource demands at different stages, and thus require elasticity, i.e., automatic resource provisioning.
\item \textbf{Programmability}: one should be able to use simple and straightforward languages to express parallelism and compose workflows with dynamic dependency graphs. 
\item \textbf{Data dependency}: handling complex data dependencies %in the computing continuum is challenging. 
across CI is complicated. 
Abstracting data movements between resources is crucial to simplify development. 
\item \textbf{Dynamic execution}: a workflow may be dynamically changing based on the prior computation results and its functions may execute dynamically on various resources based on resource availability and status.
\item \textbf{Performance}: 
given the heterogeneous and dynamic (resource availability may vary) nature of modern CI, tasks require to be scheduled to appropriate resources with high performance.
\end{enumerate}
%\end{itemize}
Existing systems only partially satisfy these requirements and in general are not designed to manage computation across federated CI. 
We are thus motivated to develop \name{}. %, a parallel programming framework for managing computation in these environments. 
\name{} leverages \funcxname{} as the main execution backend.
\funcxname{} is a federated function serving fabric and supports executing function tasks on arbitrary computing resources in a FaaS manner. 
\name{} leverages two main components in \funcxname{}: \textbf{endpoint} and \textbf{client}.

An \textbf{endpoint} represents a computing resource %on a resource
that can execute tasks using the FaaS model, i.e., the endpoint can elastically launch multiple \emph{worker} processes (or containers) and assign a task to a worker to be performed. 
One can deploy the endpoint software on any computer and integrate it into the execution fabric as an available resource.

The \funcxname{} \textbf{client} provides a secure means to interact with the cloud-hosted \funcxname{} web service to submit function tasks to specific endpoints, track task states, and retrieve task results. 
When a function task is submitted to \name{}, \name{} employs internally the \funcxname{} client to dispatch the task to the remote endpoint and retrieve the results.

We leverage \funcxname{} as the execution backend for two important reasons. First, \funcxname{} allows one to easily add/remove a resource %to/from the computing continuum through 
via its flexible endpoint software. Second, the FaaS model can inherently satisfy requirements 1--3: fine-grain function management, portability, and elasticity.
%provides several benefits such as fine-grain function management, portability, and elasticity. 
However, \funcxname{} offers only an independent task execution interface. \emph{There are still research challenges (e.g., programmability, data, dynamic execution, and performance) toward a simple programming model across a resilient, federated resource pool.}
In this paper, we discuss how we design \name{} to satisfy the requirements and solve the challenges of adapting the FaaS model for programming federated scientific workflows.

\section{Programming with \name{}}\label{sec:programming}
%\yifei{Review5: \autoref{sec:programming} and \autoref{sec:arch} does not showcase sufficiently how each requirement in \autoref{sec:requirements} is addressed.}

To satisfy the programmability requirement, \name{} adopts a Python-based programming model and allows users to compose workflows with dynamic dependency graphs.

\subsection{Programming Model}

\textbf{Functions and tasks:}
A \emph{function} can be a pure Python function or a function call to other software. Each function to be executed on remote computing resources must be decorated with \texttt{@\decorator{}}.
A \emph{task} represents an invocation request of a function. 
%While users can invoke a function in the standard Pythonic way, 
The invocation to a decorated function does not return results immediately, but instead returns a \texttt{Future} object, indicating that the function instance %corresponding to this future object 
is being executed asynchronously.
A function may accept Python objects or \texttt{Future} objects of other tasks as input arguments.
This mechanism is commonly used in existing Python-based systems such as Dask~\cite{rocklin2015dask}, Parsl~\cite{parsl}, and Ray~\cite{moritz2018ray}. 
%s aligned with the \texttt{concurrent.futures} interface used i

\textbf{Data:}
A function may be invoked with Python objects and files as input arguments. Python objects are directly serialized when passed as input arguments. There is a hard limit on the size (10 MB) of a Python object that can be passed among resources in \funcxname{}.
Any object with a size larger than the limit must be serialized and invoked as a \texttt{RemoteFile} object.

\name{} enables users to read and write files via the \texttt{RemoteFile} object in a function. When a task needs files located on a remote resource, \name{} supports transferring the files via two mechanisms: Globus~\cite{chard2014efficient} and rsync. The \texttt{RemoteFile} object includes two subclasses: \texttt{GlobusFile} and \texttt{RsyncFile} (more details in \autoref{sec:data}). We will use \texttt{GlobusFile} below to illustrate the programming model. 

% \begin{lstlisting}[label=lst:example_function, style=myPython, caption={An example function with GlobusFile. The function is to compute the fingerprint of a molecule in SMILES format.}, captionpos=b]
\begin{lstlisting}[label=lst:example_function, style=myPython, caption={An example function with GlobusFile. The function is to compute the fingerprint of a molecule in SMILES format.}, captionpos=b]    
@function
def compute_fingerprint(GlobusFile: mol_file):
    from rdkit import *
    import GlobusFile
    
    mol_path = mol_file.get_remote_file_path()
    molecule = open(mol_path).readline()
    fp = AllChem.GetMorganFingerprint(
        Chem.MolFromSmiles(molecule), 2)
        
    out_file = GlobusFile.create("fp.txt")
    out_path = out_file.get_remote_file_path()
    open(out_path, 'w').write(fp)
    return out_file
\end{lstlisting}

Listing \ref{lst:example_function} shows a \texttt{compute\_fingerprint} function that accepts a \texttt{mol\_file} parameter of type \texttt{GlobusFile}.
The path of the remote file can be retrieved via the \texttt{get\_remote\_file\_path()} method and further read/write operations on the file can be done via Python's built-in I/O libraries. One needs to call the \texttt{GlobusFile.create} method to create a new file on the remote compute resource, which returns a \texttt{GlobusFile} object that can be recognized and managed by \name{}.

\subsection{Dynamic Task Graph}
\name{} represents a workflow as a directed acyclic graph (DAG), where each node indicates a task and each edge indicates a dependency between two tasks. \name{} allows passing \texttt{future} objects as function arguments to construct task graphs. 
Specifically, when a \texttt{future} $F$ is passed to a task $T$, \name{} adds an edge from the task that is associated with $F$ to $T$. A task can execute only when all its dependencies are ready.
Through \texttt{future} passing, \name{} supports constructing a task graph \emph{dynamically}, which means that the graph can change during execution time.

\subsection{Configuration}
Each \funcxname{} endpoint is assigned a universally unique identifier (UUID) when it is deployed. 
\name{} provides a \texttt{Config} interface that allows one to specify endpoints as computing resources by their UUIDs, as shown in Listing \ref{lst:config}. 
One can also configure other parameters such as scheduling strategy, the maximum number of retries, and file transfer type.
The \texttt{Config} interface is separated from the programming interface. In other words, one can write a workflow once and deploy it on different sets of endpoints by simply updating the UUIDs specified in the workflow configuration, enabling \emph{write once, run anywhere}.

\begin{lstlisting}[label=lst:config, style=myPython,caption={An example of the \texttt{Config} interface.}, captionpos=b,]      
config = Config(
    executors=[
        Executor(label="Cluster1",
            endpoint="6156af-...-54e93"),
        Executor(label="Cluster2", 
            endpoint="9c2344-...-7ff98"),
        ],
    scheduling_strategy="LOCALITY",
    max_transfer_retries=3,
    file_transfer_type="Globus")
\end{lstlisting}

\section{Architecture and Implementation}\label{sec:arch}
\begin{figure}[h]
  \vspace{-0.1in}
  \centering
  \includegraphics[width=0.85\linewidth]{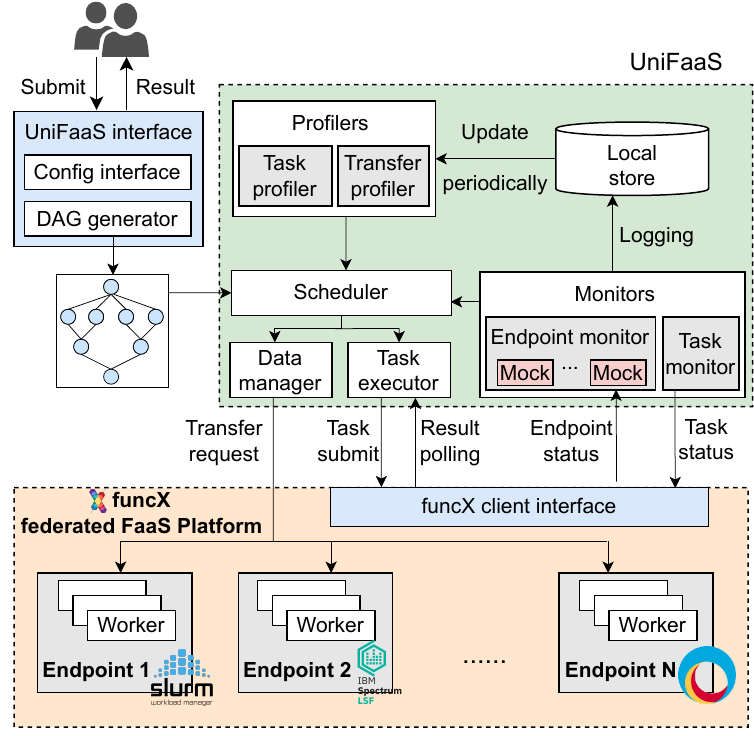}
      \vspace{-0.1in}
  \caption{\name{} architecture.}
    \vspace{-0.2in}
  \label{fig:arch}
\end{figure}

\name{} comprises five system components: monitors, profilers, scheduler, data manager, and task executor, as shown in \autoref{fig:arch}. All components are extensible to any appropriate alternatives. In this section, we describe the design of each component and discuss how \name{} can satisfy the requirements presented in \autoref{sec:requirements}.

\subsection{Overview}
% In the computing continuum,
The performance of tasks at different stages of a federated workflow may vary significantly due to the heterogeneity of both tasks and hardware.
Fortunately, decomposing a workflow into functions enables fine-grained task performance prediction.
Scientific workflows are often run repeatedly on similar sets of computers and have predictable performance~\cite{shu2021bootstrapping}.
 These characteristics together allow us to exploit an observe-predict-decide approach to improve the performance of federated workflows.
 % in the computing continuum. 
We briefly describe the execution flow of a workflow with \name{} as follows:

\begin{enumerate}[leftmargin=0.2in] 
\item Deploy \funcxname{} endpoint software on the accessible resources and supply the endpoint UUIDs in the \texttt{Config}.
\item Create the functions, decorate them with \texttt{@\decorator{}}, and express task graphs. The DAG generator analyzes the task dependencies and constructs a DAG.
\item The profilers load task characteristics from the local database (if any), build performance models, and predict execution time and data transfer time for tasks when needed by the scheduler.
\item The scheduler creates a schedule for tasks based on the information provided by the profilers and monitors.
\item Once the scheduling decisions are made, the data manager performs data transfers in advance when possible, and the task executor submits tasks to endpoints.
\item When a task runs, the task monitor tracks the characteristics and logs them into the local database after the task completes. The profilers are updated accordingly based on the latest runs.
\end{enumerate}

\subsection{Monitors}\label{sec:monitor}

\textbf{Task monitor:} This component is responsible for monitoring task execution information, such as task states (e.g., running, completed, failed, etc.), task characteristics (e.g., CPU utilization) and task completion times on various computers. The monitored information is streamed into a local database and the profilers.
%With the intuition that a user may run her/his workflow repeatedly~\cite{deelman2009workflows}. 
The local database can be treated as historical knowledge---\name{} allows a user to start a workflow by loading an existing database so that the profilers can pre-build performance models for the scheduler.
However, the task execution status may vary over time, resulting in inaccurate scheduling decisions. Therefore, real-time task information is also fed into the profilers to update the performance models.

\textbf{Endpoint monitor:}
\name{}'s scheduler requires endpoint information, including hardware configuration and real-time endpoint status (e.g., available and pending tasks), to make proper scheduling decisions.
While \funcxname{} provides RESTful interfaces for polling endpoint status, it is only updated periodically (e.g., every minute) but not in real-time, which may lead to inaccurate scheduling decisions. Moreover, frequently polling for endpoint status updates may increase the load on the \funcxname{} services, and hence is not practical.

To address this problem, we propose a \emph{local mocking mechanism}. Specifically, the endpoint monitor creates a \emph{mock endpoint} object for each endpoint listed in the \texttt{Config} interface. A mock endpoint serves as a proxy to the corresponding genuine endpoint and has the same attributes as the genuine endpoint, including hardware information, task queues, the number of busy and idle workers, etc. When initializing a mock endpoint, the endpoint monitor communicates with the \funcxname{} service to retrieve initial information. When submitting a task to a genuine endpoint, a mock task is pushed into the task queue of the mock endpoint and the number of idle workers is decreased. Similarly, the mock task is popped out of the queue after the task is completed. 
Meanwhile, to ensure the accuracy of the mock information, the endpoint monitor synchronizes the mock objects with the \funcxname{} service periodically. 
While polling-based approaches may have varying latency depending on the network latency, the overhead of this local mocking mechanism is negligible, which enables the scheduler to obtain real-time endpoint status for accurate scheduling decisions.

\subsection{Profilers}\label{sec:profilers}
\name{} uses an execution profiler and a transfer profiler to predict the execution and transfer time. These profilers are deployed on separate threads in the \name{} client. They periodically learn or update models based on real-time monitoring information and historical data from prior runs and do not interfere with the main scheduling loop.

\textbf{Execution profiler:} 
When \name{} is initialized, the execution profiler trains an initial performance model \emph{for each function} based on historical data.
When a workflow runs, the profiler updates the models periodically if there is new task information collected from the monitor. 
We apply the random forest regression algorithm~\cite{pham2017predicting,singh2017machine} as the default execution performance model. The model takes the input size, number of cores, CPU frequency, and RAM size of the endpoint to run on as inputs, and estimates the execution time and output data size.
We note that \name{}'s modular design means that users can easily extend it to other appropriate performance models such as XGBoost~\cite{chen2016xgboost} and Bayesian linear regression~\cite{bader2022lotaru}.

\textbf{Transfer profiler:}
Data transfer time is primarily determined by the data size and the network conditions between endpoints. Prior work~\cite{liu2017explaining} shows that the data transfer time is relatively predictable across federated CI.
In the absence of recent data, the transfer profiler can send probing file transfers to measure the network bandwidth between endpoints when \name{} is initialized.
The current implementation uses a polynomial regression model that uses bandwidth, data size, and the maximum number of concurrent transfers (set by the data manager in \autoref{sec:data}) to predict the data transfer time.

\subsection{Scheduler} \label{sec:scheduler}

The scheduler maps workflow tasks to heterogeneous endpoints, with the objective of minimizing the workflow completion time (i.e., makespan), which has been demonstrated to be NP-hard~\cite{ullman1975np}. 
As aforementioned, the scheduler needs to consider that both the workflow DAGs and resource capacity may vary during execution in cross-facility computing, which brings a unique challenge on how to make scheduling decisions at the most appropriate time.
Resource capacity in this paper represents the number of workers in an endpoint and each function is mapped to a worker. The resources allocated (e.g., CPUs, GPUs, memory, and disks) per worker is configurable on the \funcxname{} endpoint. The scheduler can consider various resource dimensions when making decisions.

In cross-facility computing, data staging may incur significant delay, thus it is often desirable to make task scheduling decisions as early as possible (e.g., immediately after tasks are submitted), allowing overlapping data staging and computation to reduce workflow makespan.
However, early scheduling decisions may result in poor performance in the case of dynamic workflows or resource capacity. 
To address this challenge, we propose three scheduling algorithms, capacity-aware, locality-aware, and heterogeneity-aware, to support various scenarios, from static to dynamic workflow DAGs and resource capacity.

\textbf{Capacity-aware scheduling} (\mytt{Capacity} in short): 
\mytt{Capacity} assigns workflow tasks to endpoints based on the capacity of each endpoint, i.e., the number of tasks scheduled to an endpoint is proportional to its total computing capacity. 
Assume we have a set of $N$ endpoints $\mathbb{EP} = \{e_1, e_2, ..., e_N\}$ and the capacity of the endpoints are $\mathbb{C} = \{c_1, c_2, ..., c_N\}$ (measured by the number of workers on an endpoint).
Suppose we have a set of $M$ tasks. 
The number of tasks scheduled to the endpoint $e_i$ can be computed by:
\begin{equation} \label{eq:jr}
    M_i =  M * \frac{c_i}{ \sum_{i=1}^{N} c_i}.
\end{equation}

After determining $M_i, i \in [1, N]$, the scheduler searches the candidate tasks for each endpoint in a depth-first search (DFS) order, with the intention to consider data locality and allocate tasks on the same path to the same endpoint, reducing data transferred across endpoints.

\begin{figure}[h]
  \vspace{-0.1in}
  \centering
  \includegraphics[width=\linewidth]{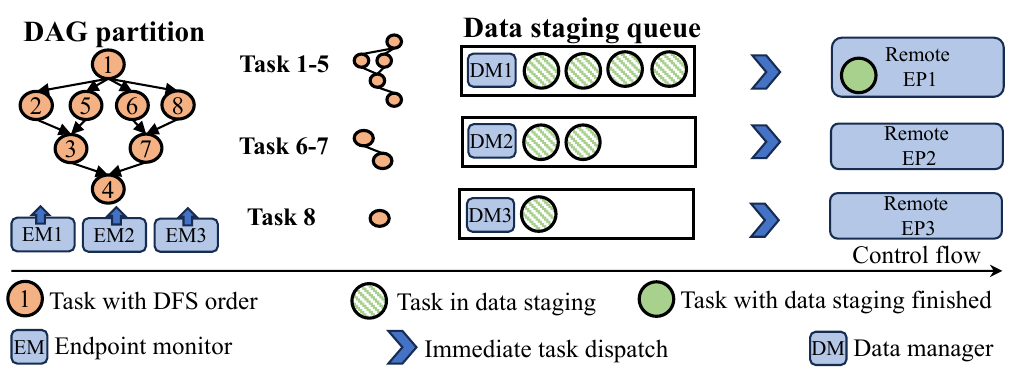}
  \vspace{-0.2in}
  \caption{An illustration of \mytt{Capacity}. EPs 1-3 have 5, 2, and 1 workers, respectively. According to the capacity and DFS order, tasks 1--5, tasks 6--7, and task 8 are assigned to endpoint 1, 2, and 3, respectively.}
  \label{fig:capacity}
\end{figure}

After partitioning the DAG based on endpoint capacity, tasks are placed into the corresponding data staging queues between endpoint pairs. Upon the completion of data staging, they are immediately dispatched to remote endpoints, as shown in \autoref{fig:capacity}.

Note that the scheduling decisions of \mytt{Capacity} are generated \emph{offline}, i.e., immediately after a workflow DAG is submitted and formed. Therefore, \mytt{Capacity} is most suitable for workflows that have static DAGs and run on endpoints with similar hardware performance and static resource capacity. 

\begin{figure}[h]
  \vspace{-0.1in}
  \centering
  \includegraphics[width=\linewidth]{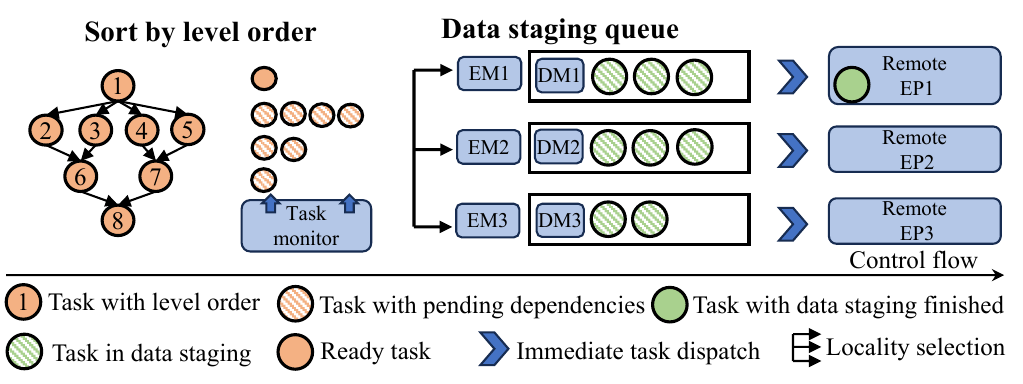}
   \vspace{-0.2in}  \caption{An illustration of \mytt{Locality}. When an endpoint monitor detects idle resources, it performs locality selection for the next ready task and immediately dispatches the task to the target endpoints.}
\vspace{-0.1in}  \label{fig:locality}
\end{figure}

\textbf{Locality-aware scheduling} (\mytt{Locality} in short):
Unlike \mytt{Capacity}, \mytt{Locality} only assigns tasks to an endpoint when there are available resources on the endpoint. Specifically, When assigning a task, \mytt{Locality} examines the data distribution for all dependencies of the task on all the endpoints. Based on the data distribution, it computes the amount of data transferred if placed on a specific endpoint and selects the endpoint that leads to the least amount of transfer, denoted as \emph{locality selection} in \autoref{fig:locality}. Upon completion of the data staging, tasks are immediately dispatched to remote endpoints.

While \mytt{Locality} is similar to \mytt{Capacity} in that they both intend to reduce data transfers across federated CI, \mytt{Locality} produces real-time scheduling decisions whenever there are idles sources and only considers current task states (e.g., data distribution and priority). 
Therefore, \mytt{Locality} is applicable to workflows with dynamic DAGs, as well as workflows that run on endpoints with dynamic resource capacity.

\begin{figure}[h]
  \vspace{-0.1in}
  \centering
  \includegraphics[width=\linewidth]{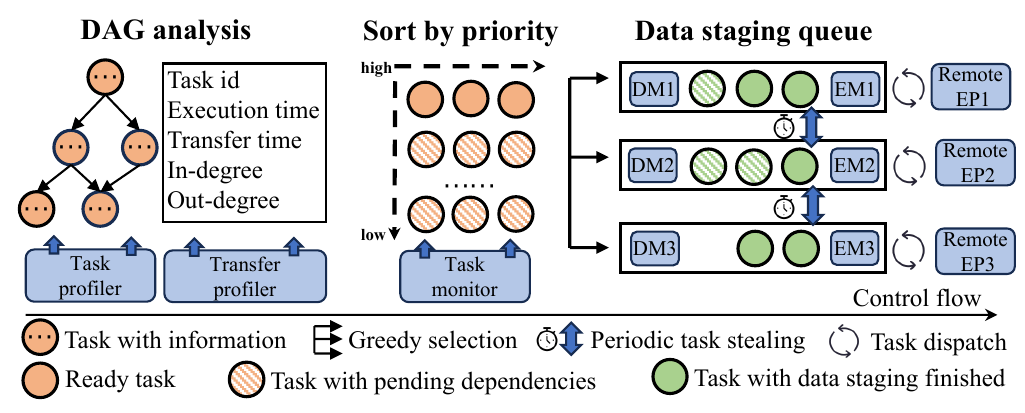}
        \vspace{-0.2in}
  \caption{An illustration of \mytt{DHA}. \mytt{DHA} involves prioritization and endpoint selection. Tasks are not dispatched until the target endpoint has idle resources.}
  \label{fig:dheft}
\end{figure}

\textbf{Dynamic heterogeneity-aware scheduling} (\mytt{DHA} in short):

\mytt{DHA} assumes that all task information and data transfer information is known a priori, via user input or from the profilers.
\mytt{DHA} involves two stages when scheduling, task prioritization and endpoint selection. \mytt{DHA} first calculates the priority of all tasks and then selects an appropriate endpoint for every task in order based on the priority, as shown in \autoref{fig:dheft}.

In detail, the priority of a task $t_i$ is recursively computed based on the following equation:

\begin{equation} \label{eq:priority}
\text{priority}(t_i) = \overline{d_i} + \overline{w_i} + \max_{t_j \in succ(t_i)}\text{priority}(t_j),
\end{equation}
where $\overline{d_i}$ denotes the average data staging time of task $t_i$ over all the endpoints, $\overline{w_i}$ is the average execution time of task $t_i$ over all the endpoints, and $succ(t_i)$ is the set of immediate successors (if there is any) of task $t_i$. 
The priority calculation is inspired by the heterogeneous earliest finish time (HEFT)~\cite{topcuoglu2002performance} algorithm. % , and demonstrates that tasks with a longer completion time are prioritized for endpoint selection. 
The recursive calculation ensures that predecessor nodes are assigned to endpoints ahead of successor nodes.

Submitted tasks are scheduled based on the priority. Once all the dependencies of a task are complete (i.e., ready task), \mytt{DHA} selects the target endpoint in a heterogeneity-aware manner that minimizes the completion time. 
After endpoint selection, \mytt{DHA} implements a \emph{delay scheduling} mechanism to delay the task dispatch to the target endpoint.
Specifically, the endpoint selection allows the task to instantiate the data staging immediately whenever a dependency of the task is complete.
However, the task dispatch is delayed until the target endpoint has idle resources. 
In other words, this mechanism allows tasks with data staging completed to wait in the \name{} client queue.

To adapt to dynamic resource capacity during execution, we propose a \emph{re-scheduling} mechanism: whenever the resource capacity changes, \mytt{DHA} periodically recomputes the scheduling decisions of pending tasks (including those with data staging completed) considering several factors (e.g., data movement cost and execution time on different clusters, and available resources) and performs task stealing if necessary.
With the \emph{delay} mechanism, the pool of tasks that can be re-scheduled is expanded. The optimization goal of the re-scheduling is to maximize the utilization of dynamic resources while reducing the completion time of the corresponding tasks.

Unlike \mytt{Capacity} and \mytt{Locality} that either makes offline or real-time scheduling decisions, 
\mytt{DHA} is a hybrid algorithm between them, i.e., it generates scheduling decisions offline but only dispatches tasks until there are idle resources.
The delay and re-scheduling mechanism are key enablers for \mytt{DHA} to support dynamic workflow DAGs and dynamic resource capacity.
\mytt{DHA} is applicable to any scenario but requires workflows to have full knowledge (e.g., graph structure, network bandwidth, and task characteristics) to have optimal performance.

\begin{table}[htbp]
  \vspace{-0.1in}
  \caption{Summary of the scheduling algorithms.}
  \centering
  \label{tab:schedule_summary}
  \begin{tabular}{{p{0.20\textwidth} p{0.07\textwidth} p{0.07\textwidth}  p{0.05\textwidth}}}
    \toprule
    % \textbf{Name}  &\textbf{Scheduling type}& \textbf{Dynamic DAG supported} & \textbf{Dynamic resource capacity supported} \\ \hline
    % \mytt{Capacity} & Offline & Static  capacity with no knowledge\\ \hline
    % \mytt{Locality}  & Real-time & Dynamic capacity with no knowledge\\ \hline
    % \mytt{DHA} & Hybrid & Static and dynamic capacity with full knowledge\\
    & \mytt{Capacity} & \mytt{Locality} & \mytt{DHA} \\ \hline
    \textbf{Scheduling type}& Offline & Real-time & Hybrid \\ 
    \textbf{Dynamic DAG supported}& \xmark &  \cmark  & \cmark\\
    \textbf{Dynamic resource supported} & \xmark & \cmark & \cmark \\
    \textbf{Knowledge required} & \xmark & \xmark & \cmark\\
  \bottomrule
\end{tabular}
\vspace{-0.1in}
\end{table}

\textbf{Summary:}
We summarize the features of the three scheduling algorithms in \autoref{tab:schedule_summary}.
The \name{} scheduler is designed to be extensible and configurable. Users may extend the scheduler simply to adapt to their use cases.

\subsection{Data Manager}\label{sec:data}
Workflows often involve complex inter-task data dependencies. While \name{} can automatically pass small objects among tasks (via \texttt{futures}), there may exist large files to stage across federated CI.
Some tools support programmatic data transfers but require real-time authentication of computers and pre-knowledge of specific locations, which misaligns with the federated workflows---a task may run on various resources and the location is unknown when programming.

\name{} resolves this via a data manager that i) provides a shim layer with \texttt{RemoteFile} and \texttt{RemoteDirectory} objects, which allows users to wrap data and supports interfaces to perform read/write operations similar to regular files and directories;
ii) stages data \emph{transparently} for users using the built-in transfer mechanisms, when a task with data dependencies is scheduled; 
and iii) monitors the progress of data transfers and retries failed transfers.

Currently, the data manager supports Globus~\cite{chard2014efficient} and rsync, two widely used mechanisms for data transfers across different resources.
The data manager also enables concurrent data transfers, and the number of threads can be configured based on the connection limit of resources or transfer mechanisms.

\subsection{Task Executor}
\name{} leverages \funcxname{} as the backend to run tasks in a FaaS manner.
Users must therefore deploy \funcxname{} endpoints on desired resources to run their tasks.
%which allows a resource to execute tasks in a FaaS manner.
The task executor asynchronously submits tasks and polls results via the \funcxname{} service. The task executor wraps the task with data staging finished as a \funcxname{} task, submits the task to the corresponding endpoint via the \funcxname{} client, and records the task’s returned future.
The task executor has a separate thread to poll the results of all pending tasks via the \funcxname{} client. 
Upon completion of a task (i.e., when its result is retrieved), 
the task executor updates the corresponding future and also streams the task execution status such as the execution time and location into the task monitor as described in \autoref{sec:monitor}.

\subsection{Fault Tolerance}
\name{} implements several fault tolerance mechanisms.

\textbf{Data transfer retry:}
Data transfers may fail due to network conditions, especially for large data volumes.  \name{} will retry failed transfers several times (configurable). If all retries fail, the corresponding tasks will be marked as failed. 

\textbf{Task reassignment:}
A task may fail for various systematic reasons, e.g., data transmission failure, endpoint disconnection, or incorrect runtime environment. 
For a failed task, \name{} attempts to execute the task again on an endpoint according to the scheduler. If it fails again, 
\name{} reassigns it to the endpoint with the highest success rate based on prior runs. If it fails on all endpoints, \name{} returns an error message.

\subsection{Optimizations}
\name{} applies several performance optimizations to ensure the efficient and robust execution of federated workflows.

\textbf{Multi-endpoint elasticity:}
%A workflow may have different levels of parallelism and 
The resource requirements of a workflow often vary at different stages. 
% \name{} implements multi-endpoint elasticity to scale connected endpoints in advance based on the characteristics of workflows.
Each \funcxname{} endpoint itself can dynamically scale: spawning more workers when there are more tasks than workers and killing idle workers when there is no incoming task for a certain period of time. 
However, each endpoint does not have a global view of workflows and may make suboptimal scaling decisions.
With a full view of workflows, \name{} can perform multi-endpoint scaling in advance based on the characteristics of workflows.

\name{} implements a \texttt{Scaling} interface which allows users to implement their own multi-endpoint scaling logic (e.g., preferences for certain endpoints).
The default multi-endpoint scaling strategy is straightforward: if the number of pending tasks in a workflow is more than the number of workers, \name{} scales out the workers on all the endpoints; the scale-in logic is left to the endpoints to decide, as each endpoint can scale in if there are idle resources (which may indicate that the endpoint is less preferred by the \name{} scheduler). 
Such an approach that \emph{scales out aggressively but scales in conservatively} works well in most use cases, since killing idle resources is generally easier than requesting resources on federated CI.

\textbf{Batching:}
Workflows may be composed of thousands of tasks.
To amortize the communication and computation costs for managing tasks and endpoints across various components,
\name{} implements batching mechanisms at several levels, including task submission, result retrieval, endpoint status polling, and performance prediction, whenever possible.

\subsection{Implementation}\label{sec:implementation}
We implement a prototype of \name{} in Python, which is based on \parslname{}~\cite{parsl}, a parallel programming library widely used in science. 
\name{} is open source on GitHub.\footnote{https://github.com/SUSTech-HPCLab/UniFaaS}

\section{Evaluation}\label{sec:evaluation}
We evaluate \name{}'s performance in terms of several system metrics including latency, scalability, elasticity, and scheduler overhead.
% We also use a real scientific workflow to study the characteristics of different scheduling strategies and demonstrate \name{}'s effectiveness.

\subsection{Testbeds}

% anonymized version
To evaluate \name{}'s ability to manage scientific workflows across federated CI, we deploy \funcxname{} endpoints on the following heterogeneous clusters, and submit workflows via \name{} on a local workstation. The hardware of these clusters and computers is listed in \autoref{tab:hardware}.

\begin{itemize}[leftmargin=0.2in] 
\item \textbf{\taiyi{}} is a 2.5-petaflops supercomputer that was in the TOP500 list until recently. %June 2022. % of June 2022
\item \textbf{\qiming{}} is a 0.3-petaflops academic supercomputer.
\item \textbf{\csecluster{}} is a department cluster used primarily for teaching and research. 
\item \textbf{\lab} is a local compute cluster. 
%\item \textbf{\pc{}} serves as both an edge resource to run tasks and the client of \name{}.
\end{itemize}

\begin{table}[htbp]
  \centering
  \footnotesize
    \vspace{-0.1in}
  \caption{Hardware of the heterogeneous testbed.}
  %\vspace{-0.1in}
  \label{tab:hardware}
  %\begin{tabular}{p{0.08\textwidth} p{0.23\textwidth} p{0.04\textwidth} p{0.03\textwidth}}
  \begin{tabular}{p{0.09\textwidth} p{0.235\textwidth} p{0.04\textwidth} p{0.03\textwidth}}
    \hline
    \textbf{Name} & \textbf{CPU} & \textbf{RAM (GB)} & \textbf{\# nodes} \\ \hline
    % \taiyi{} & 2*Intel Xeon Gold 6148@2.4 GHz & 192 & 815 \\ \hline
    % \qiming{} & 2*Intel Xeon E5-2690@2.6GHz & 64 & 230 \\ \hline
    % \csecluster{} & 2*Intel Xeon Platinum 8260@2.4GHz & 770 & 26 \\ \hline
    % \lab{}  & 2*Intel Xeon Gold 5320@2.2GHz & 128 & 2\\ \hline
    % \pc{} (client) &  Intel Core i5-9400@2.9Ghz & 16 & 1\\ \hline
    \taiyi{} & 2*Xeon Gold 6148@2.4GHz & 192 & 815 \\ 
    \qiming{} & 2*Xeon E5-2690@2.6GHz & 64 & 230 \\ 
    \csecluster{} & 2*Xeon Platinum 8260@2.4GHz& 770 & 26 \\
    \lab{}  & 2*Xeon Gold 5320@2.2GHz & 128 & 2\\ 
    \pc{} &  Core i5-9400@2.9Ghz & 16 & 1\\ \hline
  \end{tabular}
  \vspace{-0.1in}
\end{table}

\subsection{Latency}\label{sec:latency}
\begin{figure}[h]
    \vspace{-0.2in}
  \centering
  \includegraphics[width=0.85\linewidth]{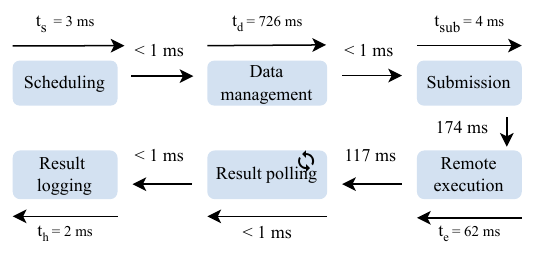}
    \vspace{-0.2in}
  \caption{\name{} latency breakdown.}\label{fig:latency}
  \vspace{-0.1in}
\end{figure}

We measure the latency incurred by each component in \name{}, %including the predictor, scheduler, data manager, and remote execution. 
by running a ``hello world'' task with a 1~MB 
input file to transfer, and timing the latency across each component. The endpoint is deployed on \qiming{} and the average of 20 runs is reported in \autoref{fig:latency}. 
The task takes around \num{1087} ms to execute. 
Each \name{} component results in only minimal latency. For example, the profilers predict job characteristics and transfer time within \num{2}~ms (included in the scheduling). The local mocking overhead, included in the submission stage, is \num{0.08}~ms.
The majority of the latency is from unavoidable data transfer, as well as task dispatching to the remote endpoint and result polling that are highly relevant to the network delay. Nonetheless, the execution times of data analysis tasks in scientific workflows often vary from minutes to hours, in which case this level of overhead is acceptable.

\subsection{Scalability}\label{sec:exp_scalability}
We evaluate the strong and weak scaling of \name{}.
Strong scaling measures performance for a fixed number of tasks, as the number of workers increases;
weak scaling measures performance for increasing numbers of workers when the average number of tasks \emph{per worker} is fixed.
It has been previously demonstrated that a single \funcxname{} endpoint can scale up to \num{130000} workers~\cite{li2022funcx,chard2020funcx}; hence, we focus here on evaluating \name{} scalability when deploying tasks across multiple endpoints---more endpoints lead to higher task scheduling and submission overheads.
In this experiment, each endpoint has 24 workers and all endpoints are deployed on \qiming{}.
% All the endpoints are warm and ready to serve tasks, which means no queue time is considered.
We create two types of tasks: 1~s and 5~s compute-intensive CPU stress (i.e., while loop) tasks.

\begin{figure}[h]
    \vspace{-0.1in}
  \centering
  \includegraphics[width=\linewidth]{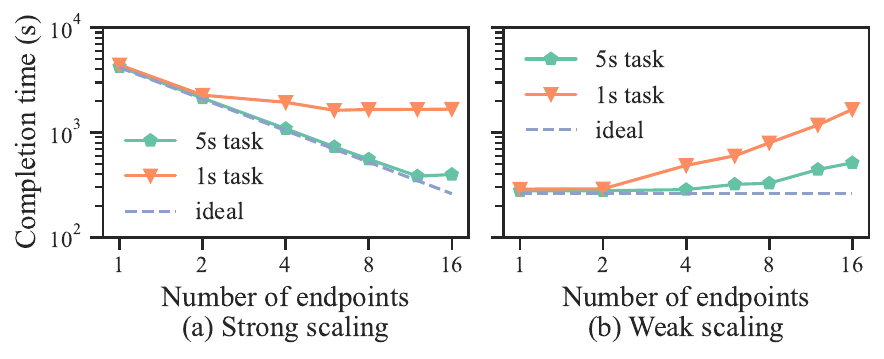}
    \vspace{-0.2in}
  \caption{Strong and weak scaling of \name{}.}\label{fig:scaling}
      \vspace{-0.2in}
\end{figure}

\autoref{fig:scaling} shows the strong and weak scaling performance from 1 to 16 endpoints.
In the strong scaling case, we measure the performance for a) \num{100000} $\times$ 
1~s tasks and b) \num{20000} $\times$ 5~s tasks. In the weak scaling case, each worker runs, on average, either a) 260 $\times$ 1~s tasks or b) 52 $\times$ 5~s tasks, yielding the same workloads in total for strong and weak scaling on 16 endpoints. We note that 16 endpoints is a sufficient number for many of our scientific use cases, but not a limit of \name{}.
The results show that the scalability for 5~s tasks is close to the ideal for up to 12 endpoints; with yet longer-duration tasks, we would expect to see good scaling for yet larger numbers of endpoints.
The completion time keeps decreasing until six endpoints for 1~s tasks and 12 endpoints for 5~s tasks. 
The performance of \num{100000} $\times$ 1~s tasks is worse than that of \num{20000} $\times$ 5~s tasks. This is primarily because a larger number of 1~s tasks suffer from higher network latency and scheduling overheads, causing worse nonlinear scaling.

\subsection{Multi-endpoint Elasticity}\label{sec:multi-endpoint}
To demonstrate the multi-endpoint elasticity of \name{}, we deployed three endpoints on \qiming{} (EP1), \csecluster{} (EP2), and \lab{} (EP3), respectively. Each endpoint can scale up and down in terms of nodes and each node has 20 workers, whereas the maximum number of workers of EP1, EP2, and EP3 is set to 100, 40, and 20, respectively. 
We create three types of compute-intensive stress tasks: 30~s tasks (task1 on EP1), 15~s tasks (task2 on EP2), and 10~s tasks (task3 on EP3). Each endpoint runs a distinct task duration since we want to show that each endpoint can scale independently.

\begin{figure}[h]
    \vspace{-0.1in}
  \centering
  \includegraphics[width=\linewidth]{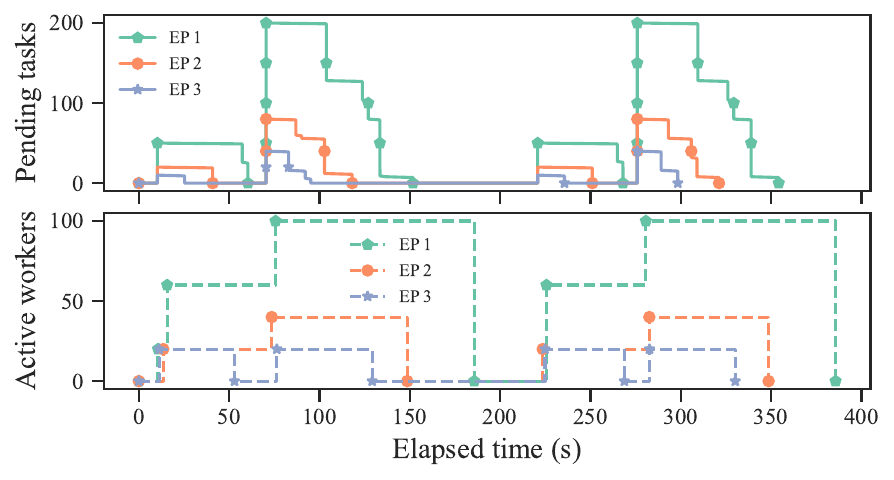}
  \vspace{-0.2in}
  \caption{Number of tasks and workers over time. Top: number of pending tasks. Bottom: number of active workers.}\label{fig:elasticity}
      \vspace{-0.1in}
\end{figure}

\autoref{fig:elasticity} shows the number of pending tasks and active workers versus time. 
At $t$ = 10, we run 50 $\times$ task1, 20 $\times$ task2, and 10 $\times$ task3. Consequently, EP1 scales up to 60 workers, while EP2 and EP3 scale up to 20 workers.
At around $t$ = 50, since EP3 has been idle for more than 30 seconds (configured maximum idle interval), EP3 returns all the workers.
At $t$ = 70, we run 200 $\times$ task1, 80 $\times$ task2, and 40 $\times$ task3 on the corresponding endpoints. This time all the endpoints scale up to the maximum number of workers (i.e., 100, 40, and 20).
After all the tasks are complete, each endpoint scales down to zero workers. We repeat the above process twice and observe that each endpoint can scale up and down promptly and independently, as expected.
It is worth mentioning that the performance of elasticity in practice is subject to the queuing delays of batch schedulers. % across federated CI.

\subsection{Scheduler Overhead}\label{sec:scheduling_overhead}
\begin{figure}[h]
  \centering
    \vspace{-0.1in}
  \includegraphics[width=0.7\linewidth]{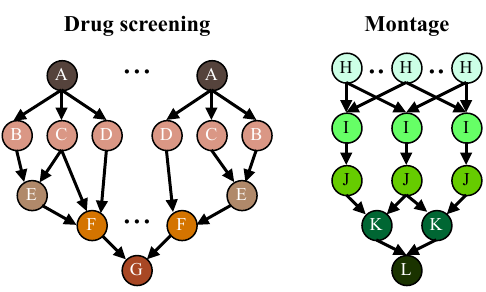}
  \caption{DAG structures of the drug screening and the montage workflows. Each character represents a distinct task type. 
  The drug screening workflow consists of \num{24001} functions. The total computation time is \num{1447} hours with an average of 220 seconds per task. The total size of the input, intermediate, and output data is 480.64~GB. The montage workflow consists of \num{11340} functions. The total computation time is \num{108}~hours with an average of 6.4~seconds per task. The total size of the input, intermediate, and output data is 673.49~GB.
  }\label{fig:workflow}

\vspace{-0.1in}
\end{figure}

We measure the scheduler overhead (including the time to predict task characteristics if needed) when scheduling a drug screening workflow in \autoref{fig:workflow}.  
Note that this overhead experiment is conducted on the \pc{} and better performance may be achievable with a more powerful server.
\autoref{tab:scheduler} shows the overhead per task of different algorithms. 
We see that all the algorithms have only a modest overhead. 
\mytt{DHA} involves predicting task characteristics and prioritizing the tasks in the DAG, resulting in a slight increase in the overhead.

\begin{table}[htbp]
  \caption{Overhead of different algorithms.}
   \vspace{-0.1in}
  \centering
  \label{tab:scheduler}
  \begin{tabular}{cc}
    \toprule
     \textbf{Scheduling algorithm} & \textbf{Overhead} (s) \\ \hline
    % Default & \num{7.74e-05} & Dynamic \\
        \mytt{Capacity} & \num{1.72e-04}   \\
    \mytt{Locality} & \num{3.00e-03}   \\
    %Heterogeneity-aware & \num{2.76e-02} & Static \\
    \mytt{DHA} & \num{3.46e-03}   \\
  \bottomrule
\end{tabular}
\vspace{-0.1in}
\end{table}

\section{Case Studies}
 We use two open-source workflows, a drug screening workflow~\cite{babuji2020targeting} and a montage workflow~\cite{berriman2004montage}, to study the characteristics of different scheduling strategies and evaluate \name{}'s ability to manage workflows across multiple resources. 

\subsection{Static resource capacity}\label{sec:case_static}

In this experiment, we use two workflows % DAG of this workflow is 
shown in \autoref{fig:workflow}. 
When executing the drug screening workflow, we deploy 2000 workers (50 nodes) on \taiyi{} (EP1), 384 workers (16 nodes) on \qiming{} (EP2), 48 workers (2 nodes) on \csecluster{} (EP3), and 52 workers (2 nodes) on \lab{} (EP4). When executing the montage workflow, we deploy 120 workers (4 nodes) on EP1, 240 workers (10 nodes) on EP2, 48 workers (2 nodes) on EP3, and 52 workers (2 nodes) on EP4. All workers are launched before the experiment, and the number of workers is \emph{static} during this experiment. For \mytt{DHA},
we assume full knowledge can be retrieved from the profilers.

\begin{table}[htbp]
  \vspace{-0.1in}
  \caption{Results for static resource capacity.} 
  \centering
  \label{tab:makespan}
  \begin{tabular}{p{0.08\textwidth} p{0.15\textwidth} p{0.07\textwidth} p{0.07\textwidth}}
    \toprule
    \textbf{Workflow} & \textbf{Experiment} & \textbf{Makespan} (s) &\textbf{Transfer size} (GB)\\ \hline
    \textbf{Drug}& \mytt{Capacity}   & \num{3240}  &  \num{4.86}\\
    &\mytt{Locality}  & \num{3882} & \num{53.46} \\
    &\mytt{DHA}  & \num{2898} & \num{44.94} \\
    &\textbf{Baseline}: Only \taiyi{} & \num{3763} & \num{0} \\\hline
    \textbf{Montage} & \mytt{Capacity}   & \num{1027}  &  \num{2.57}\\
    &\mytt{Locality}  & \num{1055} & \num{13.35} \\
    &\mytt{DHA}  & \num{909} & \num{18.27} \\
    &\textbf{Baseline}: Only \qiming{}  & \num{1994} & \num{0} \\

    \bottomrule
\end{tabular}
\vspace{-0.1in}
\end{table}

\textbf{Analysis:} 
\autoref{tab:makespan} shows the makespan of the workflows under different scheduling algorithms.
The makespan is defined as the completion time of the workflow, including scheduling overhead, polling latency, etc. 
For both workflows, \mytt{DHA} outperforms \mytt{Capacity} and \mytt{Locality} in terms of makespan, since \mytt{DHA} can leverage knowledge such as DAG structure and task characteristics. \mytt{Capacity} results in the smallest data movement because it is designed for static DAGs and can use certain knowledge like DAG structures to schedule. \mytt{Locality} operates without any prior knowledge, thereby minimizing data transfer size to the best extent possible.

\begin{figure}[h]
  \centering
  \vspace{-0.1in}
  \includegraphics[width=\linewidth]{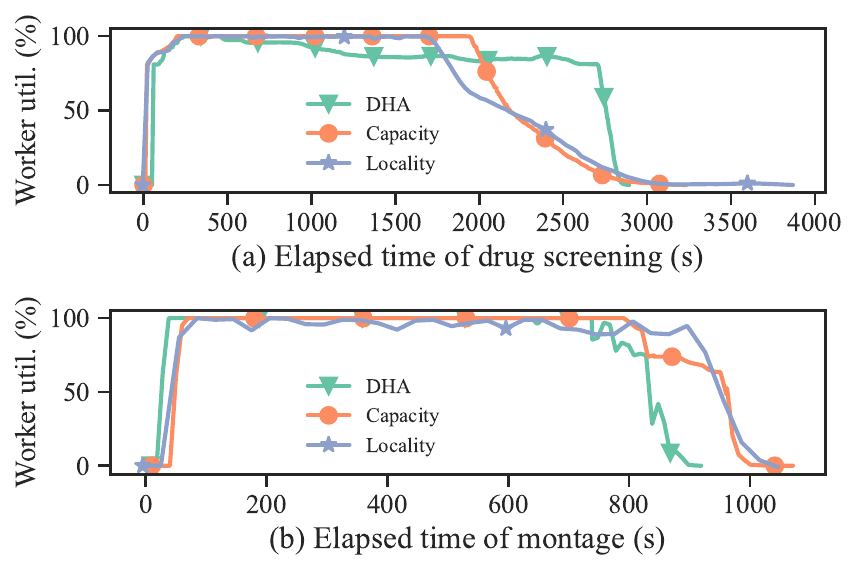}
  \vspace{-0.2in}
  \caption{Worker utilization over time under static resource capacity.}\label{fig:all_busy_workers}
  \vspace{-0.1in}
\end{figure}

\begin{figure}[h]
  \vspace{-0.1in}
  \centering
  \includegraphics[width=\linewidth]{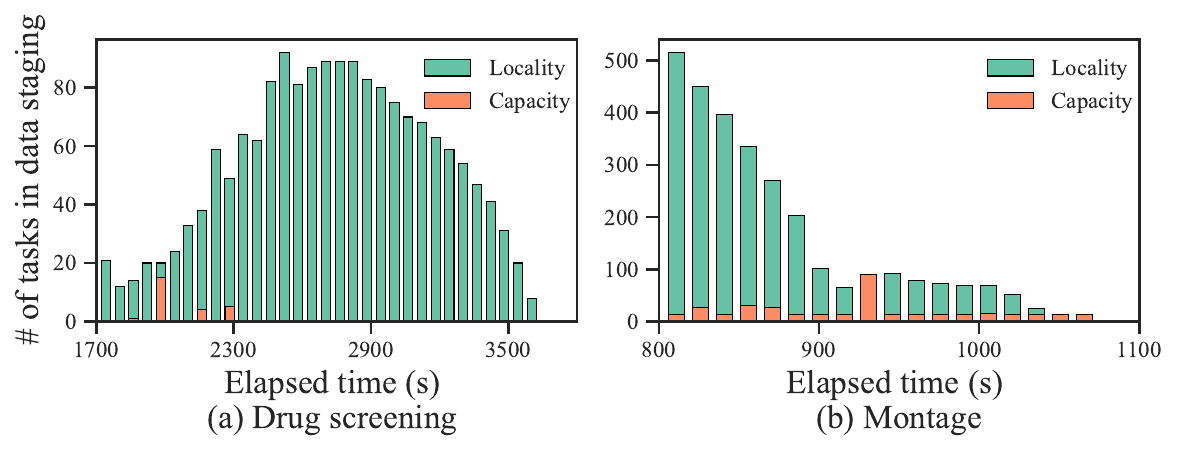}
      \vspace{-0.3in}
  \caption{\mytt{Locality} results in more tasks in data staging state compared with \mytt{Capacity}.}\label{fig:explain_locality}
    \vspace{-0.1in}
\end{figure}

\begin{figure}[h]
  \vspace{-0.1in}
  \centering
  \includegraphics[width=\linewidth]{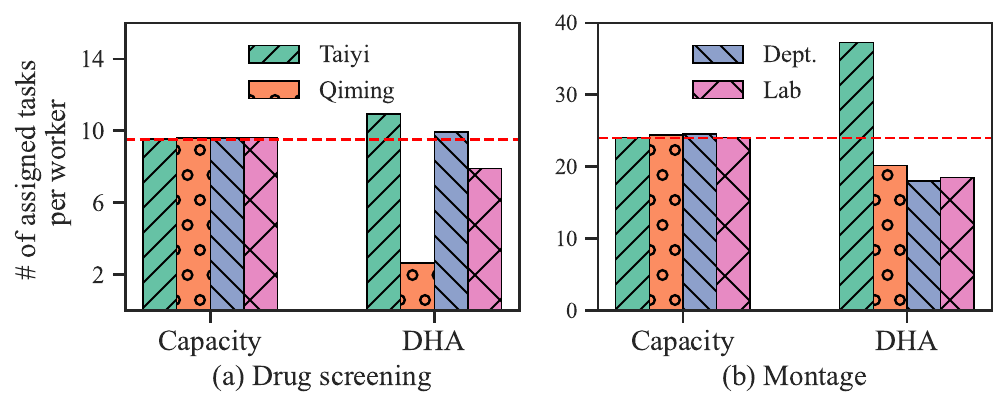}
      \vspace{-0.3in}
  \caption{Workload distribution of \mytt{Capacity} and \mytt{DHA}. \mytt{DHA} prefer \taiyi{}, a higher performance cluster.}\label{fig:explain}
    \vspace{-0.1in}
\end{figure}

To further analyze the reason for the performance variance, we plot in \autoref{fig:all_busy_workers} the worker utilization of different scheduling algorithms. 
For both workflows, \mytt{DHA} achieves consistent high worker utilization. The worker utilization of both \mytt{Locality} and \mytt{Capacity} gradually decreases, exhibiting a long-tail pattern.
%From $t$ = \num{1700}, the worker utilization of both \mytt{Locality} and \mytt{Capacity} gradually decreases, exhibiting a long-tail pattern. 
However, the root causes of these declines in worker utilization are different. 
The makespan of \mytt{Locality} is severely impacted by the data staging, since \mytt{Locality} makes scheduling decisions in real-time and cannot hide the data staging delays, resulting in a longer makespan.
This is proven by \autoref{fig:explain_locality}, which shows the number of tasks in data staging over time.
\mytt{Capacity} makes scheduling decisions offline and hence the data staging can be done immediately after each task's dependencies are completed, overlapping the data staging with the computation.
\autoref{fig:explain} demonstrates that \mytt{Capacity} can evenly distribute tasks to endpoints based on the number of workers. \mytt{DHA} is heterogeneity-aware and tends to allocate more tasks to \taiyi{} with more advanced hardware.

To demonstrate the potential of deploying workflows across federated CI with \name{}, we compare the three scheduling algorithms with a baseline, which runs on \emph{only} \taiyi{} (2000 workers) for the drug screening workflow and \emph{only} \qiming{} (240 workers) for the montage workflow. As shown in \autoref{tab:makespan}, the makespan is improved by up to 22.99\% (54.41\%) with an additional 19.48\% (47.83\%) of the workers while executing the drug screening (montage) workflow.

The results demonstrate that \emph{\name{} can effectively deploy workflows across federated CI}, even though cross-facility computing may incur a significant amount of data transfers.

\subsection{Dynamic resource capacity}\label{sec:dynamic}
\begin{figure}[h]
 \vspace{-0.1in}
  \centering
  \includegraphics[width=\linewidth]{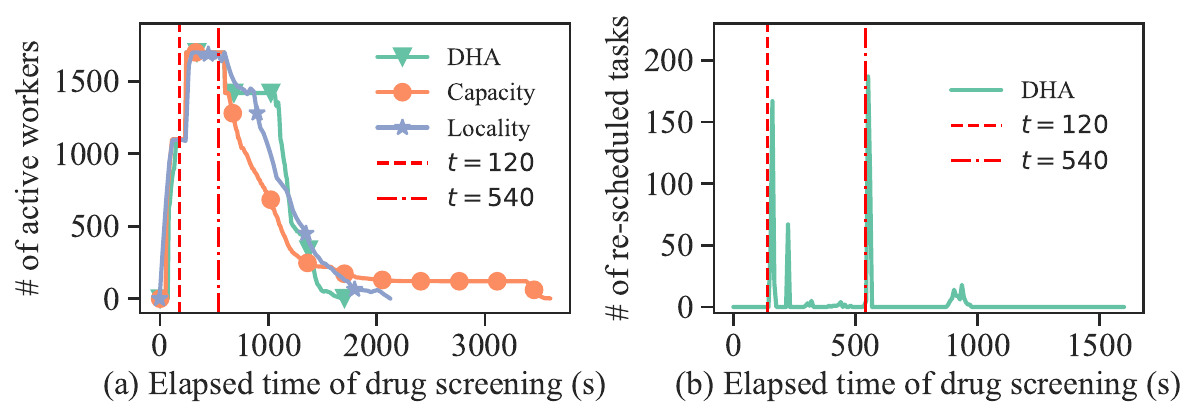}
  \vspace{-0.2in}
  \caption{Running the drug screening workflow under dynamic capacity. Initially, 400, 600, 48, and 52 workers are deployed on EP1, EP2, EP3 and EP4 respectively. At $t=120$, EP2's worker count increases by 600. At $t=540$, EP1's worker count decreases by 280.}\label{fig:dynamic_drug}
  \vspace{-0.1in}
\end{figure}

\begin{figure}[h]
 \vspace{-0.1in}
  \centering
  \includegraphics[width=\linewidth]{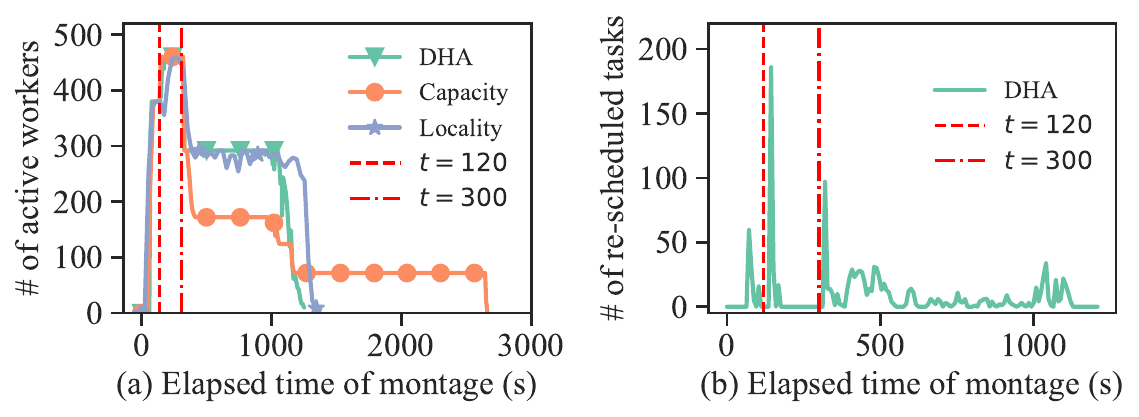}
  \vspace{-0.2in}
  \caption{Running the montage workflow under dynamic capacity. Initially, 40, 240, 48, and 52 workers are deployed on EP1, EP2, EP3 and EP4 respectively. At $t=120$, EP1's worker count increases by 80. At $t=300$, EP2's worker count decreases by 168.}\label{fig:dynamic_montage}
  \vspace{-0.1in}
\end{figure}

We study the effectiveness of \mytt{DHA} under dynamic resource capacity.
We execute the drug screening workflow with \num{12001} functions and the montage workflow with \num{11340} functions.
The captions of \autoref{fig:dynamic_drug} and \autoref{fig:dynamic_montage} show how the amount of resources varies over time (in terms of the number of workers) on the endpoints for the two workflows.

\begin{table}[htbp]
\vspace{-0.1in}
  \caption{Results for dynamic resource capacity.}
  \centering
  \label{tab:fluctuating_capacity_makespan}
  \begin{tabular}{p{0.08\textwidth} p{0.15\textwidth} p{0.07\textwidth} p{0.07\textwidth}}
    \toprule
    \textbf{Workflow} & \textbf{Experiment} & \textbf{Makespan} (s) &\textbf{Transfer size} (GB)\\ \hline
    \textbf{Drug}& \mytt{Capacity}   & \num{3610}  &  \num{3.26}\\
    \textbf{screening}&\mytt{Locality}  & \num{2130} & \num{43.61} \\
    &\mytt{DHA}  & \num{1666} & \num{33.01} \\
    &\mytt{DHA} without re-sched. & \num{2183} & \num{39.47} \\\hline
    \textbf{Montage} & \mytt{Capacity}   & \num{2671}  &  \num{2.48}\\
    &\mytt{Locality}  & \num{1360} & \num{14.18} \\
    &\mytt{DHA}  & \num{1257} & \num{31.05} \\
    &\mytt{DHA} without re-sched. & \num{1868} & \num{29.62} \\

    \bottomrule
\end{tabular}
\vspace{-0.1in}
\end{table}

\textbf{Analysis:} 
\autoref{tab:fluctuating_capacity_makespan} shows the makespan of the workflows with dynamic resource capacity for different scheduling algorithms. In both workflows, \texttt{DHA} attains the lowest makespan because the re-scheduling mechanism can promptly respond to resource variations and effectively balance the workload across endpoints. As a result, \mytt{DHA} improves the makespan by up to 32\% compared to \mytt{DHA} without re-scheduling.
\texttt{Locality} reduces the makespan by more than 41\% when compared with \texttt{Capacity} because the real-time nature of \texttt{Locality} enables it to dynamically adapt based on the current state of resources, resulting in more efficient utilization of available capacity, while operates as an offline scheduler.
\autoref{fig:dynamic_drug}(a) and \autoref{fig:dynamic_montage}(a) imply that \mytt{Capacity} fails to balance the tasks to more capable endpoints, resulting in long-tail latency due to the bottleneck endpoints. \autoref{fig:dynamic_drug}(b) and \autoref{fig:dynamic_montage}(b) show that \mytt{DHA} can quickly re-schedule tasks when there is resource variation.

\section{Lessons Learned}\label{sec:discussion}
We discuss our experiences with the case studies using \name{} and traditional approaches.

\textbf{Traditional approaches versus \name{}:}
%When experimenting with \name{}, 
We attempted to compare the performance of \name{} with traditional systems such as
Pegasus~\cite{deelman2015pegasus} in the evaluation.
However, it was challenging %to conduct a fair comparison 
since traditional systems are not designed for those requirements or do not run workflows in a manner like \name{} does.
For example, Pegasus is coupled with HTCondor, which needs to open a port on each computing resource and establish a direct connection to the submit host to listen for incoming jobs. 
However, this is rarely allowed by HPC clusters and requires privileged access.

Further, traditional systems launch tasks of workflows at the job level and are not suitable for workflows with dynamic DAGs and dynamic resource capacity. 
\name{} differs in that it allows one to programmatically compose a workflow and map the workflow to a resilient resource pool at the function level.
\name{} enables construction of workflows to run across federated cyberinfrastructure using Python, while traditional systems often require domain-specific languages or specifications to describe workflows.

In summary, \name{} differs from traditional approaches in terms of both \emph{programmability} and \emph{task management}.
Such a FaaS-based approach creates new opportunities for considering workflows in which small schedulable units can be flexibly placed across \emph{dynamic} federated computers.

\textbf{Applicability to federated workflows:}
We utilized two supercomputers, \taiyi{} and \qiming{}, in our evaluation. 
%Based on the discussions with the administrators,
\taiyi{}, boasting new generation hardware, %and a TOP500 ranking, 
usually has longer queue times than \qiming{}. 
In general, users run workflows on either \taiyi{} or \qiming{}, since it is a burden to manage workflows across two heterogeneous clusters using traditional approaches.
%Remarkably,
\name{} provides the ability to easily explore tradeoffs between hardware performance and queue time, using \taiyi{} and \qiming{}, as well as the \csecluster{} and \lab. % any resource that one can access.
Based on our observations, the tasks of the workflow use \qiming{} preferably when \taiyi{} is busy, and prefer \taiyi{} when its resources are available. %(with multi-endpoint elasticity turned on).

\textbf{Programmability and Portability:} 
We developed and debugged the drug screening and montage workflows in \autoref{sec:evaluation} locally, and finally ran them across several computing resources. These workflows were originally designed to run a single cluster. We made them executable across distributed CI with minimal effort using \name{}: i)  decorate each stage as functions; ii) replace the original file I/O operations with \texttt{RemoteFile} object; and iii) deploy the resource pool and specify the endpoints' UUID. During the development, the main computation code remained unchanged. 
%\name{} enabled us to transition from the local development to the multi-endpoint computations with minimal effort, i.e., deploying new endpoints on those new resources. %and adding them in the \texttt{Config} interface.
Additionally, \qiming{} was updated from the PBS to the LSF scheduler during our experiments.
With \name{}, we transitioned the workflow to the updated \qiming{} by merely replacing the \funcxname{} endpoint with one configured for LSF.
These experiences demonstrate that \name{} simplifies the utilization of diverse resources.

\section{Related Work}\label{sec:related}

\textbf{FaaS.} 
FaaS platforms are widely offered by most cloud providers~\cite{amazonlambda,AzureFunctions,googlecloudfunctions}. 
%such as AWS Lambda~\cite{amazonlambda}, Azure Functions~\cite{AzureFunctions}, and Google Cloud Functions~\cite{googlecloudfunctions}. 
There are also many open-source FaaS frameworks (e.g., %OpenFaaS~\cite{openfaas}, 
OpenWhisk~\cite{openwhisk}, KNIX~\cite{knix}, and DFaaS~\cite{ciavotta2021dfaas}) that allow users to deploy on-premise and for different scenarios (e.g., IoT). The success of the FaaS model in clouds motivates us to adopt and extend the FaaS model for modern science workflows.
We thus build \name{} upon \funcxname{}~\cite{li2022funcx,chard2020funcx}, a specialized FaaS platform for federated CI. 

Several papers~\cite{malawski2020serverless,jonas2017occupy,shankar2020numpywren,carver2020wukong,mahgoub2021sonic} focus on migrating DAG-like applications to use the FaaS model. For instance, PyWren~\cite{jonas2017occupy} and NumPyWren~\cite{shankar2020numpywren} leverage FaaS for specific types of applications such as MapReduce and linear algebra; Wukong~\cite{carver2020wukong} is a serverless parallel programming framework that relies on AWS Lambda. These works are tied to centralized (cloud-hosted) FaaS platforms (e.g., Lambda) or are limited to just one endpoint. Our novelty lies in the use of the FaaS model as a way to distribute computation across federated CI. We focus on the unique challenges of such federated environments (e.g., scheduling and data transfer). 
% We have further identified the key requirements and have explored several techniques to meet these requirements.

\textbf{Workflow management systems.}
While there are many workflow management systems developed~\cite{deelman2015pegasus,wilde2011swift,jain2015fireworks,di2017nextflow,luigi,airflow}, none aim to address the needs of fine-grain task execution across federated CI specifically, to the best of our knowledge.
For instance, Pegasus~\cite{deelman2015pegasus} similarly aims to bridge distributed and diverse CI. However, Pegasus relies on a static DAG model and requires HTCondor~\cite{thain2005distributed} as the broker to interact with different cluster schedulers.
\name{} instead leverages a Python-based dynamic DAG model and the flexibility of \funcxname{} endpoints allows one to simply run on arbitrary resources.
%Further, \name{} natively supports orchestrating functions into workflows, providing fine-grained task-level management.
Python parallel frameworks (e.g., Dask~\cite{rocklin2015dask}, Parsl~\cite{parsl}, Ray~\cite{moritz2018ray}) support construction of parallel programs with Python functions and simple deployment on various types of clusters (e.g., supercomputers and clouds). 
While \name{} uses a similar programming interface, these frameworks are primarily designed for deployment on a single cluster. % but not programming in the computing continuum.

\textbf{Workflow performance modeling and scheduling.}
Estimating runtimes and other task characteristics of workflows is a well-studied area~\cite{hilman2018task,bader2022lotaru,wyatt2018prionn,silva2013toward,nadeem2017modeling,pham2017predicting,yu2022workflow}. 
\name{} relies on several performance models to predict task characteristics and transfer performance.  
%We would like to reiterate that performance modeling is not our focus. 
The profilers of \name{} are designed to be extensible to any of these models.

Workflow management systems~\cite{deelman2015pegasus,bux2015saasfee,di2017nextflow} rely on efficient scheduling algorithms to map workflows to target resources.
Many papers propose workflow scheduling algorithms for different scenarios. %(e.g., heterogeneity-aware). 
For example, prior papers~\cite{topcuoglu2002performance,dai2014synthesized,barbosa2011dynamic,kumar2021delta} focus on scheduling tasks efficiently to heterogeneous processors. BaRRS~\cite{casas2017balanced} leverages task graph partitioning and data replication to reduce the data transfers among different resources.
\name{} addresses a unique scheduling problem in the federated FaaS scenario, where workflow DAGs could be dynamic and resources may be added (or removed) during execution.
However, there are common experiences and techniques in these papers that we can draw lessons from or further integrate into \name{}. For example, the priority calculation in the \mytt{DHA} algorithm is adopted from HEFT~\cite{topcuoglu2002performance}.
% These papers are orthogonal to our work and 

\section{Conclusion}\label{sec:conclusion}
\name{} adopts a federated FaaS model that enables users to focus on \emph{what} to do in the functions and \emph{when} to invoke the functions tasks of workflows, without considering the management of execution. 
\name{} provides a unified, function-based programming interface for expressing task parallelism and composing task graphs, as well as providing transparent data management.
Internally, \name{} monitors task characteristics, creates performance models for tasks on different endpoints, and decides where to dispatch tasks to achieve high performance in large-scale, heterogeneous, and dynamic environments. We demonstrated that \name{} introduces only minimal latency overhead and that \name{} can support a workflow to deploy efficiently across up to 16 different endpoints. 
In the future, %we plan to investigate how to automatically balance the tradeoff between early scheduling and dynamic execution, without requiring users to configure manually.
we intend to incorporate additional data transfer profilers into \name{}, which will consider various factors such as communication patterns and network bandwidth.
We will investigate more comprehensive scheduling algorithms and explore the coordination of these algorithms with multi-endpoint elasticity to enhance resource utilization and performance.

\section*{ACKNOWLEDGMENT}
%We thank the anonymous reviewers for their guidance during the preparation of our camera-ready version.
This work was supported in part by the National Natural Science Foundation of China Grant No. 62202216, the Guangdong Basic and Applied Basic Research Foundation Grant No. 2023A1515010244, the Shenzhen Science and Technology Program Grant 20231121101752002, the DOE contract DE-AC02-06CH11357, and the U.S. National Science Foundation Grant 2004894. This work was also supported by Center for Computational Science and Engineering at Southern University of Science and Technology.

% \section*{References}
\newpage

\balance

\bibliographystyle{IEEEtran}
\bibliography{refs}

\end{document}